# Computational Intelligence and Neuroscience

# Image Denoising Using Sparsifying Transform Learning and Weighted Singular Values Minimization

Yanwei Zhao,[1] Ping Yang,[1] Qiu Guan,[1] Jianwei Zheng [1] and Wanliang Wang[1]

[1] Zhejiang University of Technology, HangZhou 310023, China.

Correspondence should be addressed to Jianwei Zheng; zjw@zjut.edu.cn; Wanliang Wang: wwl@zjut.edu.cn

## Abstract

In image denoising (IDN) processing, the low-rank property is usually considered as an important image prior. As a convex relaxation approximation of low rank, nuclear norm based algorithms and their variants have attracted significant attention. These algorithms can be collectively called image domain based methods, whose common drawback is the requirement of great number of iterations for some acceptable solution. Meanwhile, the sparsity of images in a certain transform domain has also been exploited in image denoising problems. Sparsity transform learning algorithms can achieve extremely fast computations as well as desirable performance. By taking both advantages of image domain and transform domain in a general framework, we propose a sparsity transform learning and weighted singular values minimization method (STLWSM) for IDN problems. The proposed method can make full use of the preponderance of both domains. For solving the non-convex cost function, we also present an efficient alternative solution for acceleration. Experimental results show that the proposed STLWSM achieves improvement both visually and quantitatively with a large margin over state-of-the-art approaches based on an alternatively single domain. It also needs much less iteration than all the image domain algorithms.

## Introduction

Noise inevitably exists in images during the process of real-world scenes acquisition by reason of physical limitations, leading to image denoising (IDN) becomes a fundamental task in image processing. The recent IDN can be categorized as data-driven and prior-driven approaches.

The data-driven methods turn to certain deep convolution neural network, such as Universal Denoising Net (UDN) [1] and Fractional Optimal Control Net [2], for the IDN problem. These CNN models, although have achieved great success provided with sufficient training samples, may not perform well in small-scale data applications. For example, one cannot obtain the acceptable network parameters on a single corrupted image, which is the case considered in this study. The aim of the prior-driven methods for image denoising is to renovate the inferior image by certain image prior or other properties, such as local smoothness, non-local similarity, low-rank structure and so forth [3-5]. More specifically, the prior-based image denoising process means to find the inherently ideal image from the degraded one by extracting the few significant factors and excluding the noisy information. It is a typical ill-posed linear inverse problem, and a widely used image degradation model can be generally formulated as [6-9]:

$$Y = HX + N, \qquad (1)$$

where $X$, $Y$ are both matrices representing the original image and the degraded one, respectively. $H$ is also a matrix denoting the non-invertible degradation operator and $N$ is the additive noise.

To cope with the ill-posed problem, the general image denoising problem can be formulated as [9, 10]:

$$\min \| HX - Y \|_F^2 \quad s.t. \ F(X), \qquad (2)$$

where $F(X)$ is regarded as the image prior knowledge, including local smoothness, non-local similarity, low-rank and sparsity, $\|\cdot\|_F$ denotes Frobenius norm. According to sparsity property, the degraded image $x$ ($x$ is the vectorization of $X$, $x \in R^n$) satisfies $x = D\kappa + e$, where $D \in R^{n \times m}$ is a synthesis over-complete dictionary, $\kappa \in R^m$ is the sparse coefficient and $e$ is an approximation term in image domain [11]. This model is called as *synthesis model*, and $\kappa$ is supposed sparse ($\|\kappa\|_0 \ll m$).

To be specific, given an image $x$, the synthesis sparse coding problem is subject to finding a sparse $\kappa$ to minimize $\| x - D\kappa \|_2^2$. Various algorithms have been proposed [12-16] to figure out this NP-hard problem. Numerous researchers have learned the synthesis dictionary and updated the non-zero coefficients simultaneously to well represent the potential high-quality image. And these methods have been demonstrated useful in image denoising. Specifically, these synthesis models typically alternate two steps: the sparse coding updating and dictionary learning. However, the practical operation of synthesis models requires some rigorous conditions, which often violate in applications.

While the synthesis model has attracted extensive attentions, the *analysis model* has also been catching notice recently [17, 18]. The analysis model considers that a noisy image $x \in R^n$ satisfies $\|\Omega x\|_0 \ll m$, where $\Omega \in R^{n \times m}$ is regarded as an analysis dictionary, since it 'analyzes' the image $x$ to a sparse form. The essence of $\Omega x$ defines the subspace to which the image belongs. And the underlying ideal image is formulated as $y = x + \xi$, with $\xi$ representing noise. The denoising problem is to find $x$ by minimizing $\| y - x \|_2^2$ subject to $\|\Omega x\|_0 \ll m$. This problem is also NP-hard and resemblant of sparse coding in the synthesis model. Approximation algorithms of learning analysis dictionary have been proposed in recent years, which similar to the synthesis case are also computationally expensive.

More recently, a generalized analysis model named transform learning model has been proposed, which follows the intuition that images are essential sparse in certain transform domain and can be expressed as: $Wx = \mu + \varepsilon$, where $W \in R^{m \times n}$ is transform matrix, $\mu \in R^m$ is sparse coefficient, and $\varepsilon$ is approximation error [19]. The distinguishing feature from the synthesis and analysis models is that approximation error $\varepsilon$ of transform learning model is in transform domain and is likely to be small. Another superiority of transform model compared to image domain model is that the former can achieve exact and extremely fast computations.

Instead of learning synthesis or analysis dictionary, the transform learning model aims at learning the transform matrix to minimize the approximation error $\varepsilon$. After getting the learned transform $W$, the original image is recovered by $W^{\dagger}\mu$, where $W^{\dagger}$ is pseudo-inverse of $W$. The transform learning model has earned great success in application of image denoising in both efficiency and effectiveness [19-22].

Nonetheless, a remaining drawback is that transform model overemphasizes transform domain but ignores the primary image domain. There is always a connection between image domain and transform domain, and this can be treated as a regularization term in image denoising.

For taking full use of the advantages both image domain and transform domain, and implementing single image denoising problem, this study focuses on sparsifying transform learning and essential sparsity property of image, and proposes a novel algorithm named Sparsifying Transform Learning and Weighted Singular Values Minimization (STLWSM). Specifically, our model simultaneously considers the sparsifying transform learning and the weighted singular values minimization of image patches.

The remainder of this paper is organized as follows. In the next section, a brief review of the transform domain and image domain for IDN is provided. In section 3, we propose our method and the efficient obtain of solution. Section 4 provides experimental results of gray images and color images. Conclusions are drawn in section 5.

## Related Works

**Transform domain for IDN**

As mentioned in the previous section, the transform model can utilize the sparsity of image in transform domain to increase efficiency. Therefore the analytical transform models such as Wavelets and discrete cosine transform (DCT) are widely used in practical application, for instance, the image compression standards JPEG2000. As a classical and effective tool, transform models have been increasingly used in image denoising. Inspired by dictionary learning, Saiprasad et.al [20] proposed a Learning Sparisifying Transform (LST) model. In [20], for any noisy image $X \in R^{h \times l}$, it is first reformed to another resolution as $X' \in R^{p \times N}$, where each column represents a square patch of the original $X$ extracted by a sliding window. Second, a transform matrix $W \in R^{p \times p}$ is randomly initialized to formulate the transform sparse coding problem as follows:

$$\min \|WX' - \mu\|_F^2 - \lambda \lg \det |W| \quad s.t. \|\mu_i\|_0 \leq s \; \forall i \qquad (3)$$

where $\mu \in R^{p \times N}$ is sparse coefficient, $\mu_i$ is the column of $\mu$, $s$ is a constant representing the sparse magnitude. The additional regular term $\lambda \lg \det |W|$ is used to avoid a trivial solution. $\lambda$ is a balance coefficient, and $\lg \det |W|$ is the log-determinant of $W$ with base 10. Saiprasad et.al [20] solved the proposed problem by alternately updating $W$ and $\mu$, and proved the convergence. To carry forward their achievements, they further proposed a Learning Doubly Sparse Transforms (LDST) for IDN [22]. Specifically, $W' = B\Phi$ is adopted to replace the original $W$, where $B$ and $\Phi$ are both square matrices with the same size. $B$ is a transform constrained to be sparse, and $\Phi$ is an analysis transform with an efficient implementation.

They use doubly sparse transform model in image denoising and get faster and better results than unstructured transforms. And then, Wen et.al [19, 21] proposed a Structured Overcomplete Sparsifying Transform Learning (SOSTL) model. The main feature different from aforementioned transform models is that Wen et.al cluster image patches and learn diverse $W$ for corresponding patch groups. This process can be formulated as the following:

$$\min \sum_{k=1}^{K} \left\{ \sum_{i \in C_k} \|W_k X'_i - \mu_i\|_2^2 + \lambda_k Q(W_k) \right\} \quad s.t. \|\mu_i\|_0 \leq s \ \forall i, \{C_k\} \in G \tag{4}$$

where $Q(W) = -\log|\det W| + \|W\|_F^2$ is a regular term to prevent trivial solutions. $\{C_k\}$ indicates the specific class of image $X'$, $K$ is the number of categories and $G$ is the set of all classes.

**Image domain for IDN**

While the transform learning models have achieved great success, in image domain, there also have been proposed various algorithms for IDN. As mentioned before, in general image denoising model, $F(X)$ is an additional regularization. The widely studied regularizations include $l_1$, $l_2$, $l_{1/2}$ norm, nuclear norm, low-rank property and so on [23-25]. Focusing on patch form instead of vector form, low-rank property has been attracting significant research interest. As a convex relaxation of low-rank matrix factorization problem (LRFM), the nuclear norm minimization (NNM) has engrossed more attention [4, 25-27]. The nuclear norm of an image $X$ is defined as $\|X\|_* = \sum_i |\sigma_i(X)|_1$, where $\sigma_i(X)$ is $i$-th singular value of $X$. However, many researchers hold that the minimization of different singular values should be separated. Gu et.al [4] proposed weight nuclear norm minimization (WNNM) for image denoising problems. The weight nuclear norm is defined as $\|X\|_{w,*} = \sum_i |w_i \sigma_i(X)|_1$, and $w = [w_1, w_2, \ldots, w_n]$ is non-negative. At this point, we can treat $F(X)$ as $F(X) = \|X\|_{w,*}$, and the denoising model is:

$$\min_X \|X - Y\|_F^2 + F(X), \quad F(X) = \|X\|_{w,*} \tag{5}$$

By taking consideration of different singular values, as well as image structure, the WNNM shows strong denoising capability. Meanwhile, Hu et.al [27] proposed Truncated Nuclear Norm Regularization (TNNR) for matrix completion. They deemed that the minimization of the smallest $\min(m, n)-r$ singular values can maintain the original matrix rank by holding the first $r$ nonzero singular values fixed. Using $F(X) = \sum_{i=r+1}^{\min(m,n)} \sigma_i(X)$, the TNNR constrained model can be written as follows:

$$\min_X \|X - Y\|_F^2 + F(X), \quad F(X) = \sum_{i=r+1}^{\min(m,n)} \sigma_i(X) \tag{6}$$

TNNR gets a better approximation to the rank function than the nuclear norm based approaches. Inspired by both WNNM and TNNR, Liu et.al [28] improved the previous algorithms by reweighting the residual error separately and minimizing the truncated nuclear norm of error matrix simultaneously (TNNR-WRE). In their work, $F(X)$ is considered as follows:

$$F(X) = \| X - Y \|_* - tr(U_r H V_r')  \tag{7}$$

where $H = X-Y$, $U$ and $V$ are left and right matrices of $H$'s singular value decomposition (SVD) respectively, and $r$ is the truncation parameter. TNNR-WRE further achieves higher accuracy than TNNR.

From the above, the nuclear norm based algorithms usually can get considerable results because of the essential low-rank property in image domain. For taking both advantages of transform domain and image domain in IDN, a Sparsifying Transform Learning and Weighted Singular Values Minimization (STLWSM) method is proposed. In contrast to LST, LDST, SOLST, WNNM, TNNR and TNNR-WRE, the proposed STLWSM jointly takes consideration of sparsity in transform domain and low-rank in image domain. The main results of our work can be enumerated as follows:

(i) We propose a general framework of image process in both transform domain and image domain, which combines the sparsifying transform learning of image patches and the low-rank property of the original image.

(ii) As image patches can take advantage of the non-local similarity exists inherently in image, we learn the sparsifying transform for each group of similar patches by Euclidean distance.

(iii) For solving the proposed NP-hard problem, we present an efficient alternative optimization algorithm. In practical applications, our method requires limited number of iterations, mostly less than 3, for the final solution.

(iv) We applied our model to IDN, the results show that STLWSM can achieve evident PSNR (Peak Signal to Noise Ratio) improvements over other state-of-the-art methods.

## Proposed method

In this section, we propose a general framework in both transform domain and image domain. To be clear, we take sparsifying transform learning in transform domain and weighted singular values minimization in image domain simultaneously. To solve this NP-hard problem, an efficient solution is also derived.

**Sparsifying Transform Learning and Weighted Singular Values Minimization (STLWSM)**

In light of the observations mentioned above, we first introduce a sparsifying learning transform base on image patches, and utilize the weighted singular values minimization to improve the image quality.

Given a noisy image $X \in R^{h \times l}$, nonlocal similarity is a well-known patch-based prior which means that one patch in one image has many similar patches [7-9]. Accordingly, overlapped image patches can be extracted with a sliding window in fixed step size. For each specific patch, we choose the most similar $M$ patches by Euclidean distance [4,7,19-21] for potential low-rank structure, and a matrix of $X_i' \in R^{p \times M}$ is constructed. The patch's size is $\sqrt{p} \times \sqrt{p}$, and the total number $N'$ of $X'_i$ depends on the size of the original image $X$, patch size and step size. After similar patches aggregation process, each group $X'_i$ is obtained and $X' = [X'_1, X'_2, ..., X'_{N'}] \in R^{p \times M \times N'}$. Following the idea of transform learning algorithm [19-21], with the obtained $X'_i$ and some initialized $W_i$, our preliminary model can be formulated as the following:

$$\min_{W_i} \sum_{i=1}^{N'} \|W_i X'_i - \mu_i\|_F^2 - \lambda_i Q(W_i) \quad s.t. \|\mu_i\|_0 \leq s \; \forall i \tag{8}$$

The definition of $Q(W_i)$ is the same as one in problem (4), but $\mu_i \in R^{p*M}$ is the sparse representation of $X'_i$ in transform domain, which is a matrix. Suppose the transform $W_i$ and sparse coefficient $\mu_i$ has been updated. The denoised patch can be obtained by $X''_i = W_i^\dagger \mu_i$. Obviously, $X''_i$ also has low-rank structure, hence, we utilize weighted singular values minimization to approximate the matrix. The unified denoising minimization is:

$$\min_{W_i, \mu_i} \sum_{i=1}^{N'} \|W_i X'_i - \mu_i\|_F^2 + \alpha_i \|\mu_i\|_0 + \beta_i \|W_i^\dagger \mu_i\|_{w,*} - \lambda_i Q(W_i) \tag{9}$$

where $\alpha_i$ and $\beta_i$ are regularization parameters and usually set empirically. This formulation can minimize the residual in transform domain and the rank of the recovered matrix $X''_i$ simultaneously.

**Efficient optimization of the proposed model**

In this subsection, we introduce an efficient solution for the non-convex Sparsifying Transform Learning and Weighted Singular Values Minimization problem. According to [17-20], the transform learning process is not sensitive to the initialization of $W$. As a result, with given $W$, the sub-problem of $\mu_i$ can be obtained using cheap hard-thresholding, $\hat{\mu}_i = Th_s(\mu_i)$. Here $Th_s(\cdot)$ is the hard thresholding operator. And the sub-problem of $W_i$ is as follows:

$$\min_{\mu_i} \sum_{i=1}^{N'} \|W_i X'_i - \mu_i\|_F^2 + \beta_i \|W_i^\dagger \mu_i\|_{w,*} - \lambda_i \log|\det W_i| + \lambda_i \|W_i\|_F^2$$
$$= \min tr\{W_i (X'_i X'^T_i + \lambda_i I_p) W_i^T - 2W_i X'_i \mu_i^T + \mu_i \mu_i^T\} - \lambda_i \log|\det W_i| + \min \beta_i \|W_i^\dagger \mu_i\|_{w,*} \tag{10}$$

Because of the term $\beta_i \|W_i^\dagger \mu_i\|_{w,*}$ is more like a postfix operator, we divide the updating process of $W_i$ into two parts:

$$\begin{cases} a. \; \min tr\{W_i (X'_i X'^T_i + \lambda_i I_p) W_i^T - 2W_i X'_i \mu_i^T + \mu_i \mu_i^T\} - \lambda_i \log|\det W_i| \\ b. \; \min_{\mu_i} \sum_{i=1}^{N'} \|W_i X'_i - \mu_i\|_F^2 + \min \beta_i \|W_i^\dagger \mu_i\|_{w,*} \end{cases} \tag{11}$$

*a*. The first formula is:

$$\min tr\{W_i(X'_iX'^T_i + \lambda_i I_p)W_i^T - 2W_iX'_i\mu_i^T + \mu_i\mu_i^T\} - \lambda_i \log|\det W_i| \quad (12)$$

Decomposing $X'_iX'^T_i + \lambda_i I_p$ as $Z_iZ_i^T$, $O_i = W_iZ_i$. Then $W_iX'_i\mu_i^T$ can be written as $O_iZ^{-1}X'_i\mu_i^T$. Let $O_i$ and $Z^{-1}X'_i\mu_i^T$ have full SVD of $U\Phi V^T$ and $P\Psi Q^T$ respectively. If we take consideration of their diagonal matrix only, the foregoing formula can be rewritten as:

$$\begin{aligned}
&\min tr\{W_i(X'_iX'^T_i + \lambda_i I_p)W_i^T - 2W_iX'_i\mu_i^T + \mu_i\mu_i^T\} - \lambda_i \log|\det W_i| \\
&= \min tr(O_iO_i^T) - 2tr(O_iZ^{-1}X'_i\mu_i^T) - \lambda_i \log|\det OZ^{-1}_i| \\
&= \min tr(O_iO_i^T) - 2tr(O_iZ^{-1}X'_i\mu_i^T) - \lambda_i(\log|\det O_i| - \log|\det Z^{-1}|) \\
&= \min[tr(\varphi^2) - 2\max(tr(U_i\Phi_iV_i^T \cdot P_i\Psi_iQ_i^T)) - \lambda_i(\log|\det O_i|)] \\
&\leq \min \sum_{i=1}^n \varphi_i^2 - 2\sum_{i=1}^n \varphi_i\psi_i - \lambda_i\sum_{i=1}^n \log\varphi_i
\end{aligned} \quad (13)$$

where $\log|\det Z^{-1}|$ is constant and can be omitted. The revised problem is convex for $\varphi_i$, so the optimizing solution can be found by taking partial differential with respect to $\varphi_i$ and setting the derivative to 0.

$$\begin{aligned}
0 &= \frac{\partial(\sum_{i=1}^n \varphi_i^2 - 2\sum_{i=1}^n \varphi_i\psi_i - \lambda_i\sum_{i=1}^n \log\varphi_i)}{\partial\varphi_i} \\
&= \frac{\partial(\varphi_i^2 - 2\varphi_i\psi_i - \lambda_i\log\varphi_i)}{\partial\varphi_i} \\
&= 2\varphi_i - 2\psi_i - \lambda_i \frac{1}{\varphi_i \ln 10}
\end{aligned} \quad (14)$$

Therefore, excluding the non-positive results, the solution is:

$$\varphi_i = \frac{\psi_i + \sqrt{\psi_i^2 + \frac{2\lambda_i}{\ln 10}}}{2} \quad (15)$$

To sum up, the transform update step can be computed as follows:

$$\begin{aligned}
\hat{W}_i &= \hat{O}_iZ^{-1} = U_i\hat{\Phi}_iV_i^TZ_i^{-1} \\
&= \frac{U_i}{2}(\Psi_i + (\Psi_i + 2\lambda_i I_p/\ln 10)^{\frac{1}{2}})V_i^TZ_i^{-1}
\end{aligned} \quad (16)$$

*b*. The second formula is:

$$\min_{\mu_i} \sum_{i=1}^{N'} \|W_iX'_i - \mu_i\|_F^2 + \beta_i \|W_i^\dagger \mu_i\|_{w,*} \quad (17)$$

With fixed $\hat{W}_i$ obtained in step. a, this part can be simply seen as:

$$\min_{\boldsymbol{\mu}_i} \sum_{i=1}^{N'} \| \boldsymbol{W}_i \boldsymbol{X}'_i - \boldsymbol{\mu}_i \|_F^2 + \beta_i \cdot \boldsymbol{L}_{\boldsymbol{W}_i^\dagger \boldsymbol{\mu}_i} w_i \boldsymbol{\Sigma}_{\boldsymbol{W}_i^\dagger \boldsymbol{\mu}_i} \boldsymbol{R}_{\boldsymbol{W}_i^\dagger \boldsymbol{\mu}_i} \tag{18}$$

where $\boldsymbol{L}_{\boldsymbol{W}_i^\dagger \boldsymbol{\mu}_i} w_i \boldsymbol{\Sigma}_{\boldsymbol{W}_i^\dagger \boldsymbol{\mu}_i} \boldsymbol{R}_{\boldsymbol{W}_i^\dagger \boldsymbol{\mu}_i} = \text{SVD}(\boldsymbol{W}_i^\dagger \boldsymbol{\mu}_i)$, and $\boldsymbol{W}_i^\dagger \boldsymbol{\mu}_i$ represents the denoised matrix. Following Ref. [4], a desirable weighting vector $w_i$ in image domain can be:

$$w_i = c\sqrt{M} / (\sigma_i(\boldsymbol{W}_i^\dagger \boldsymbol{\mu}_i) + \varepsilon) \tag{19}$$

where $\sigma_i(\boldsymbol{W}_i^\dagger \boldsymbol{\mu}_i)$ is $i$-th singular of $\boldsymbol{W}_i^\dagger \boldsymbol{\mu}_i$, $c$ is a positive constant, $\varepsilon = 10^{-16}$ is to avoid dividing by zero. And the second formula's optimal solution is:

$$\hat{\boldsymbol{X}}''_i = \boldsymbol{L}_{\boldsymbol{X}''_i} \boldsymbol{S}_w (\boldsymbol{\Sigma}_{\boldsymbol{X}''_i}) \boldsymbol{R}_{\boldsymbol{X}''_i}^\text{T} \tag{20}$$

where $\boldsymbol{X}''_i = \boldsymbol{W}_i^\dagger \boldsymbol{\mu}_i$ and the soft-thresholding operator $\boldsymbol{S}_w(\boldsymbol{\Sigma}_i)$ is defined as $\boldsymbol{S}_w(\boldsymbol{\Sigma}_i) = \max(\boldsymbol{\Sigma}_i - w_i, 0)$.

---

Algorithm 1  Efficient Solution of STLWSM

Input: $\boldsymbol{X} \in R^{h \times l}$ -noisy image with size $h \times l$, $p$ –patch size, $M$ –number of similar patches, initial sparsity $s_\mu$; $\alpha_i, \beta_i, \lambda_i$ - the constants.

Output: $\hat{\boldsymbol{X}} \in R^{h \times l}$ -denoised image

Initialization: $W_i$ is the DCT matrix of size $p \times p$, $N'$ is number of similar patches' group.

For iteration =1:3
Do:
For each group ($i$=1:$N'$) calculate:
1) Transform domain:
   a. Decompose the image $\boldsymbol{X}$ into patch form $\boldsymbol{X}'$.
   b. Compute $\boldsymbol{\mu}_i$ by $\boldsymbol{\mu}_i = \boldsymbol{W}_i \boldsymbol{X}'_i$
   c. Update $\boldsymbol{W}_i$ by $\hat{\boldsymbol{W}}_i = \dfrac{\boldsymbol{R}}{2}(\boldsymbol{\Sigma} + (\boldsymbol{\Sigma} + 2\lambda_i \boldsymbol{I}_p)^{\frac{1}{2}}) \boldsymbol{Q}^\text{T} \boldsymbol{L}^{-1}$
2) Image domain:
   a. Compute $\sigma_i(\boldsymbol{W}_i^\dagger \boldsymbol{\mu}_i)$ by $\text{SVD}(\boldsymbol{W}_i^\dagger \boldsymbol{\mu}_i)$
   b. Compute $w_i = c\sqrt{M} / (\sigma_i(\boldsymbol{W}_i^\dagger \boldsymbol{\mu}_i) + \varepsilon)$
   c. Compute $\hat{\boldsymbol{X}}''_i = \boldsymbol{U}_\mu \boldsymbol{S}_w(\boldsymbol{\Sigma}_\mu) \boldsymbol{V}_\mu^\text{T}$
End.
3) Image reconstruction.

---

The summary of our optimization solution is presented in Algorithm.1, where the similar patches are determined by Euclidean distance.

## Experiment Results

In this section, we choose 25, 12, 15, 10 reference images with size of 256*256 from TID2008 [29], USC-SIPI[1], Live-IQAD [30], IVC-SQDB [31] to test the image denoising effects, respectively. As we use six different noise levels to the test images in our experiments, the total number of distorted images is 372. Some representative images from USC-SIPI database are shown in Fig. 1 and Fig. 2. Four recently proposed methods, including patch-based algorithm GSR, weighted nuclear norm WNNM, sparsity learning transform scheme SOLST and sparsity transform learning and low-rank model STROLLR, are adopted as contrasts. The noisy images are obtained by additional Gaussian noise with $\sigma_n$ = 15, 20, 30, 40, 50, 75. All competing algorithms use their default settings, which has been finely tuned and deeply verified in their original publications. Since that our method is derived from both the schemes of image domain and transform domain, we set our parameters the same as the representative methods in these two domains, i.e., WNNM and SOLST, for fairness. That is, for the image denoising application, when $\sigma_n \leq 20$, $p$ is 6, $M$ is 70, $\lambda_i$ is 0.54. When $20 < \sigma_n \leq 40$, $p$ is 7, $M$ is 90, $\lambda_i$ is 0.56. When $40 < \sigma_n \leq 60$, $p$ is 8, $M$ is 120, $\lambda_i$ is 0.58. And when $\sigma_n$ is set others, $p$ is 9, $M$ is 140, $\lambda_i$ is 0.58. In addition, 6 images of 512*512 from USC-SIPI (shown in Fig.10) are used in image inpainting application. For the image inpainting application, we also follow the similar setting rule. The balance parameters $\alpha_i$ and $\beta_i$ are both set as $\alpha_i = \beta_i = 10*\|X'_i\|_F^2$. Table 1 shows the detailed parameter setting in our experiments, where the texts in bracket is used for the 512*512 images, while the plain ones are for the 256*256 images.

Table 1 Parameter setting in our experiments

| $\sigma_n (\sigma_m)$ | 15 (15%) | 20 (20%) | 30 (30%) | 40 (40%) | 50 (50%) | 75 |
|---|---|---|---|---|---|---|
| $p$ | 6 (12) | | 7 (14) | | 8 (16) | 9 |
| $M$ | 70 (200) | | 90 (260) | | 120 (300) | 140 |
| $\lambda_i$ | 0.54 (0.54) | | 0.56 (0.56) | | 0.58 (0.58) | 0.58 |
| $\alpha_i$ | | | $10*\|X'_i\|_F^2$ ($10*\|X'_i\|_F^2$) | | | |
| $\beta_i$ | | | $10*\|X'_i\|_F^2$ ($10*\|X'_i\|_F^2$) | | | |

The Peak Signal-to-Noise Ratio (PSNR) and Structural Similarity Index Measure (SSIM) are used to evaluate the quality of the denoised images. PSNR is defined by:

$$\text{PSNR} = 10*\log_{10}\frac{255}{\text{MSE}},$$

where MSE is the mean squared error between the original image and the denoised one. SSIM is defined as [30,32]:

$$\text{SSIM}(x, y) = \frac{(2\mu_x\mu_y + C_1)(2\sigma_{xy} + C_2)}{(\mu_x^2 + \mu_y^2 + C_1)(\sigma_x^2 + \sigma_y^2 + C_2)},$$

---

[1] http://sipi.usc.edu/database/

where *x* and *y* represent the original image and the denoised one, respectively, $\mu_x$ and $\mu_y$ are the mean values of *x* and *y*, $\sigma_x$ and $\sigma_y$ are variances, and $\sigma_{xy}$ is the covariance. $C_1$ and $C_2$ denote two stabilization variables.

For a thorough comparison, we list the average denoising results from all the 372 distorted images in Table 2. Also, the experimental results from all the gray images of USC-SIPI are shown in Table 3.

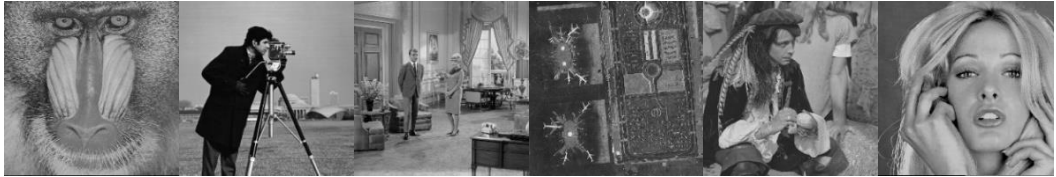
Fig. 1 Original gray images

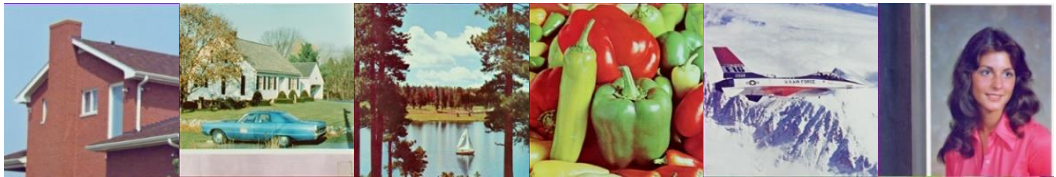
Fig. 2 Original color images

Table 2 Average denoising results with different noise level (PSNR/SSIM)

| $\sigma_n$ | GSR | WNNM | SOLST | STROLLR | STLWSM |
|---|---|---|---|---|---|
| 15 | 38.63/0.574 | 50.78/0.935 | 47.74/0.914 | 42.99/0.677 | **55.49/0.983** |
| 20 | 35.25/0.431 | 48.36/0.925 | 45.34/0.894 | 40.11/0.584 | **53.20/0.978** |
| 30 | 31.23/0.293 | 45.88/0.903 | 41.91/0.822 | 36.67/0.465 | **52.90/0.973** |
| 40 | 28.59/0.216 | 43.40/0.874 | 39.45/0.790 | 33.72/0.375 | **50.52/0.962** |
| 50 | 26.60/0.165 | 43.82/0.811 | 37.54/0.734 | 31.70/0.328 | **50.75/0.955** |
| 75 | 23.04/0.095 | 41.19/0.488 | 34.05/0.609 | 28.17/0.254 | **48.07/0.920** |

From these two tables, we can observe that among the competing algorithms, GSR also adopts the nonlocal similarity that groups image patches for low-rank structure. However, it requires too much iterations in practical applications, e.g., 100 or even up to 200 times. In contrast, WNNM needs fewer iterations, around 14, and achieves pretty good results than other 3 algorithms at average of 8.26dB for gray images. In the meantime, the proposed STLWSM needs the least iterations and achieves best performance.

SOLST and STROLLR are both transform algorithms and have hard-to-catch efficiency. STROLLR trains transform matrices for each group, while SOLST combines non-local low-rank and transform learning, they also achieved better results than STROLLR at average of 5.54dB. In Table 3, the numerical results of the proposed STLWSM are all made bold that means the best one among the five algorithms. It's evident that the proposed method has achieved visible improvement in PSNR under all kinds of noise levels at average of 13.61dB. More visual results are shown in Fig. 3, in which our method clearly outperforms all other methods.

Moreover, considering that GSR needs too much iterations and pure transform learning algorithms are extremely faster, we compare our time consummation against WNNM, and the results are shown in Fig. 4. It can be seen that our method spends much less time than WNNM, at average of 55.46%.

Table 3 Gray images de-noising results (PSNR/SSIM)

| Image | $\sigma_n$ | GSR | WNNM | SOLST | STROLLR | STLWSM |
|---|---|---|---|---|---|---|
| Baboon | 15 | 38.27/0.765 | 50.89/0.981 | 47.76/0.960 | 43.00/0.752 | **55.74/0.992** |
|  | 20 | 35.41/0.650 | 48.42/0.967 | 45.35/0.933 | 40.12/0.640 | **53.36/0.987** |
|  | 30 | 31.60/0.478 | 45.98/0.937 | 41.91/0.865 | 36.34/0.469 | **53.08/0.979** |
|  | 40 | 29.01/0.359 | 43.45/0.896 | 39.45/0.788 | 33.70/0.356 | **50.63/0.963** |
|  | 50 | 27.03/0.275 | 43.89/0.809 | 37.54/0.710 | 31.71/0.280 | **50.88/0.955** |
|  | 75 | 23.47/0.154 | 41.21/0.360 | 34.05/0.535 | 28.18/0.170 | **48.15/0.906** |
| Camera | 15 | 39.32/0.577 | 50.74/0.979 | 47.72/0.959 | 42.89/0.741 | **55.32/0.990** |
|  | 20 | 35.77/0.429 | 48.36/0.964 | 45.32/0.932 | 40.04/0.629 | **53.11/0.985** |
|  | 30 | 31.67/0.295 | 45.89/0.935 | 41.92/0.864 | 36.31/0.458 | **52.81/0.976** |
|  | 40 | 29.03/0.225 | 43.43/0.894 | 39.45/0.788 | 33.69/0.347 | **50.47/0.961** |
|  | 50 | 27.04/0.179 | 43.79/0.806 | 37.54/0.709 | 31.70/0.271 | **50.71/0.952** |
|  | 75 | 23.47/0.112 | 41.16/0.359 | 34.05/0.535 | 28.17/0.162 | **48.06/0.902** |
| Couple | 15 | 38.82/0.719 | 50.82/0.980 | 47.75/0.960 | 42.95/0.746 | **55.57/0.991** |
|  | 20 | 35.61/0.584 | 48.40/0.967 | 45.34/0.933 | 40.09/0.634 | **53.26/0.986** |
|  | 30 | 31.64/0.411 | 45.87/0.936 | 41.90/0.865 | 36.33/0.463 | **52.98/0.978** |
|  | 40 | 29.02/0.305 | 43.44/0.895 | 39.45/0.288 | 33.76/0.350 | **50.57/0.963** |
|  | 50 | 27.03/0.233 | 43.82/0.807 | 37.54/0.711 | 31.69/0.275 | **50.83/0.954** |
|  | 75 | 23.47/0.132 | 41.19/0.359 | 34.05/0.535 | 28.18/0.166 | **48.14/0.905** |
| Lax | 15 | 38.39/0.751 | 50.80/0.980 | 47.76/0.959 | 42.64/0.717 | **55.65/0.992** |
|  | 20 | 35.46/0.636 | 48.38/0.966 | 45.34/0.931 | 39.86/0.600 | **53.29/0.986** |
|  | 30 | 31.61/0.470 | 45.88/0.935 | 41.91/0.863 | 36.20/0.425 | **52.97/0.977** |
|  | 40 | 29.01/0.357 | 43.39/0.894 | 39.45/0.787 | 33.59/0.313 | **50.55/0.962** |
|  | 50 | 27.02/0.277 | 43.83/0.806 | 37.53/0.710 | 31.61/0.241 | **50.78/0.953** |
|  | 75 | 23.47/0.160 | 41.20/0.359 | 34.05/0.536 | 28.11/0.140 | **48.08/0.903** |
| Man | 15 | 38.79/0.690 | 50.74/0.979 | 47.73/0.959 | 42.88/0.739 | **55.37/0.990** |
|  | 20 | 35.61/0.552 | 48.36/0.965 | 45.33/0.931 | 40.03/0.627 | **53.13/0.985** |
|  | 30 | 31.64/0.381 | 45.89/0.935 | 41.90/0.863 | 36.29/0.454 | **52.85/0.976** |
|  | 40 | 29.02/0.279 | 43.38/0.894 | 39.45/0.786 | 33.67/0.342 | **50.48/0.961** |
|  | 50 | 27.03/0.212 | 43.77/0.806 | 37.53/0.708 | 31.67/0.267 | **50.71/0.952** |
|  | 75 | 23.47/0.119 | 41.19/0.358 | 34.05/0.533 | 28.15/0.159 | **48.05/0.903** |
| Woman1 | 15 | 39.02/0.648 | 50.81/0.996 | 47.75/0.960 | 43.08/0.759 | **55.53/0.991** |
|  | 20 | 35.68/0.499 | 48.34/0.990 | 45.34/0.933 | 40.18/0.650 | **53.22/0.986** |
|  | 30 | 31.66/0.333 | 45.87/0.936 | 41.90/0.865 | 36.39/0.479 | **52.91/0.978** |
|  | 40 | 29.02/0.240 | 43.38/0.895 | 39.45/0.788 | 33.73/0.366 | **50.52/0.962** |
|  | 50 | 27.03/0.181 | 43.81/0.807 | 37.54/0.710 | 31.71/0.289 | **50.74/0.954** |
|  | 75 | 23.47/0.102 | 41.19/0.358 | 34.05/0.534 | 28.20/0.177 | **48.06/0.904** |

Our algorithm also has good scalability, we further use RGB images in IDN, experiments results show that the proposed STLWSM still outperform than other algorithms, and specific numerical comparison are shown in Table 4. Again, Fig. 5 and Fig. 6 respectively show the visual results in terms of the average PSNR and the elapsed time, which also demonstrate our superiority against other competitors. Fig. 7 and Fig. 8 show the visual results of average SSIM comparison of gray images and color images respectively. It can be seen that our method can hold denoised image structure even with high noise rate.

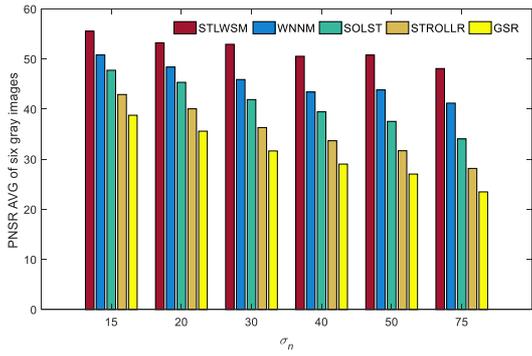

Fig. 3 PSNR AVG of gray images denoising results

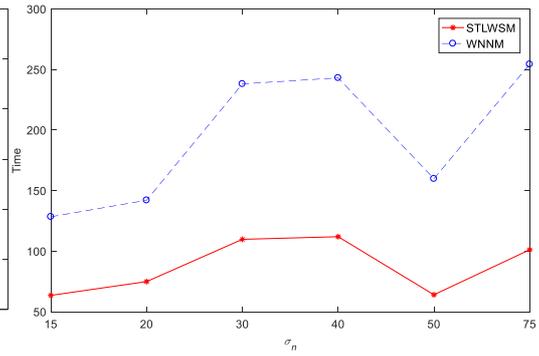

Fig. 4 Elapsed Time comparison in gray images

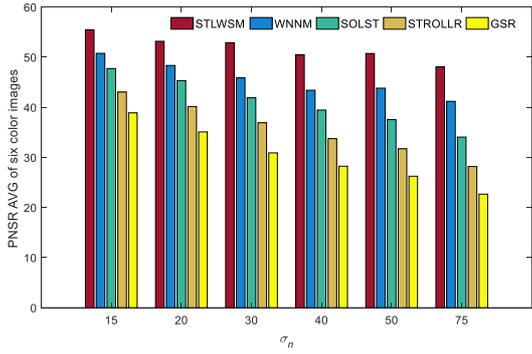

Fig. 5 PSNR AVG of color images denoising results

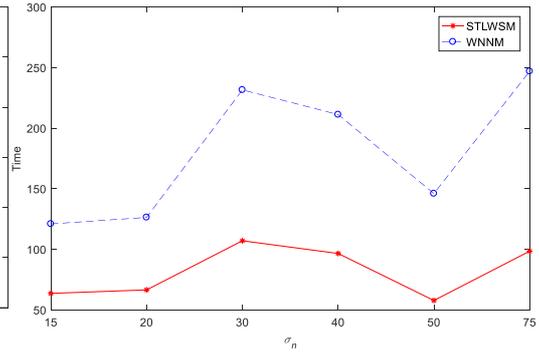

Fig. 6 Elapsed Time comparison in color images

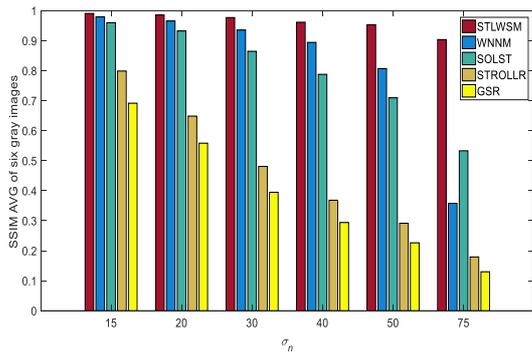

Fig. 7 SSIM AVG of gray images denoising results

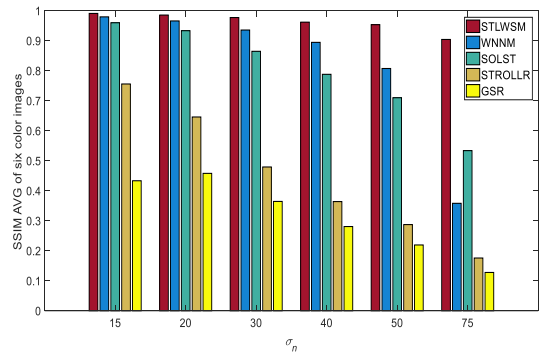

Fig. 8 SSIM AVG of color images denoising results

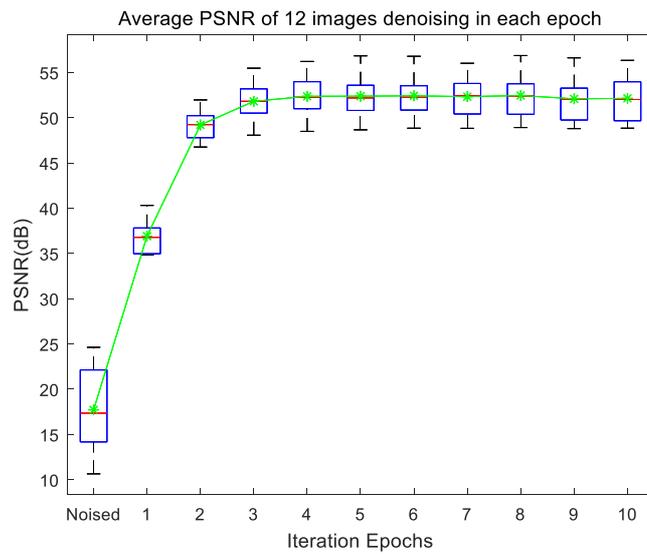

Fig. 9 Average PSNR of 12 images denoising in each epoch of different image noise levels

To detailed display the efficiency of our algorithm, we provide its generated results versus different iterations (up to 10). The experimental results are shown in Fig. 9. All 12 images' PSNR values are averaged for each noise level. The PSNR value of the original noisy images in different noise levels is shown as the starting point, where the top black line is the max value of 24.63, the bottom black line is the min value of 10.65, and the red line represents median of 17.64. And the green star is average of 17.72. Fig. 7 shows that our algorithm has a fast constringency speed and needs limited number of iterations, mostly 3, for the final solution.

Table 4 Color images de-noising results (PSNR/SSIM)

| Image | $\sigma_n$ | GSR | WNNM | SOLST | STROLLR | STLWSM |
|---|---|---|---|---|---|---|
| House | 15 | 38.89/0.507 | 50.87/0.980 | 47.76/0.960 | 43.06/0.756 | **55.73/0.992** |
|  | 20 | 35.08/0.550 | 48.43/0.967 | 45.35/0.933 | 40.14/0.645 | **53.35/0.987** |
|  | 30 | 30.90/0.448 | 45.96/0.937 | 41.91/0.865 | 36.37/0.477 | **53.04/0.978** |
|  | 40 | 28.24/0.356 | 43.40/0.896 | 39.45/0.788 | 33.73/0.364 | **50.63/0.963** |
|  | 50 | 26.24/0.268 | 43.86/0.808 | 37.54/0.710 | 31.71/0.287 | **50.86/0.955** |
|  | 75 | 22.67/0.152 | 41.19/0.359 | 34.05/0.534 | 28.19/0.176 | **48.14/0.905** |
| House2 | 15 | 38.20/0.329 | 50.78/0.980 | 47.74/0.960 | 43.19/0.770 | **55.55/0.992** |
|  | 20 | 34.87/0.319 | 48.37/0.967 | 45.33/0.933 | 40.26/0.663 | **53.25/0.986** |
|  | 30 | 30.85/0.265 | 45.88/0.937 | 41.90/0.865 | 36.44/0.466 | **52.97/0.978** |
|  | 40 | 28.22/0.211 | 43.42/0.896 | 39.44/0.789 | 33.80/0.383 | **50.57/0.963** |
|  | 50 | 26.23/0.172 | 43.82/0.809 | 37.54/0.710 | 31.76/0.278 | **50.84/0.955** |
|  | 75 | 22.67/0.109 | 41.22/0.358 | 34.05/0.533 | 28.21/0.190 | **48.15/0.907** |
| Lake | 15 | 38.10/0.484 | 50.67/0.979 | 47.71/0.959 | 42.93/0.746 | **55.29/0.990** |
|  | 20 | 34.83/0.461 | 48.27/0.965 | 45.31/0.932 | 40.08/0.635 | **53.09/0.985** |
|  | 30 | 30.84/0.381 | 45.83/0.935 | 41.89/0.864 | 36.32/0.466 | **52.82/0.977** |
|  | 40 | 28.22/0.291 | 43.38/0.895 | 39.44/0.788 | 33.71/0.354 | **50.48/0.962** |
|  | 50 | 26.23/0.226 | 43.78/0.808 | 37.54/0.710 | 31.68/0.278 | **50.72/0.954** |
|  | 75 | 22.67/0.130 | 41.22/0.359 | 34.04/0.534 | 28.16/0.169 | **48.08/0.906** |
| Pepper | 15 | 38.52/0.535 | 50.74/0.978 | 47.74/0.959 | 42.94/0.744 | **55.28/0.989** |
|  | 20 | 34.97/0.492 | 48.33/0.964 | 45.33/0.932 | 40.07/0.632 | **53.06/0.984** |
|  | 30 | 30.88/0.439 | 45.84/0.933 | 41.98/0.864 | 39.52/0.476 | **52.73/0.975** |
|  | 40 | 28.23/0.344 | 43.35/0.892 | 39.45/0.787 | 33.69/0.348 | **50.45/0.959** |
|  | 50 | 26.24/0.279 | 43.81/0.805 | 37.54/0.709 | 31.70/0.272 | **50.61/0.950** |
|  | 75 | 22.67/0.158 | 41.18/0.358 | 34.05/0.534 | 28.16/0.164 | **47.95/0.900** |
| Plane | 15 | 38.44/0.451 | 50.82/0.980 | 47.74/0.961 | 43.31/0.782 | **55.57/0.992** |
|  | 20 | 34.94/0.431 | 48.37/0.967 | 45.34/0.939 | 40.35/0.676 | **53.25/0.987** |
|  | 30 | 30.87/0.348 | 45.88/0.937 | 41.91/0.867 | 36.51/0.511 | **52.95/0.979** |
|  | 40 | 28.23/0.265 | 43.41/0.896 | 39.45/0.790 | 33.85/0.397 | **50.55/0.963** |
|  | 50 | 26.23/0.204 | 43.85/0.809 | 37.54/0.712 | 31.81/0.318 | **50.79/0.955** |
|  | 75 | 22.67/0.117 | 41.17/0.358 | 34.05/0.534 | 28.23/0.200 | **48.14/0.906** |
| Woman2 | 15 | 38.82/0.398 | 50.73/0.979 | 47.74/0,959 | 42.89/0.737 | **55.36/0.990** |
|  | 20 | 35.07/0.390 | 48.32/0.965 | 45.34/0.932 | 40.03/0.623 | **53.11/0.985** |
|  | 30 | 30.90/0.302 | 45.86/0.934 | 41.90/0.863 | 36.29/0.451 | **52.73/0.976** |
|  | 40 | 28.23/0.227 | 43.42/0.893 | 39.44/0.786 | 33.67/0.338 | **50.39/0.960** |
|  | 50 | 26.24/0.174 | 43.84/0.805 | 37.53/0.708 | 31.67/0.264 | **50.61/0.951** |
|  | 75 | 22.67/0.102 | 41.14/0.357 | 34.04/0.533 | 28.14/0.157 | **47.93/0.901** |

We also applied our method in image inpainting with 6 images in sizes of 512*512, and the degenerated images are obtained by multiplying with a random logical matrix in element-wise manner, and the missing rates are setting as $\sigma_m$={15%, 20%, 30%, 40%, 50%}. The image inpainting results are shown in Table 5. The original images are shown in Fig. 10. The results show that all methods achieve admirable inpainting results for filling in missing pixels, and the proposed STLWSM still outperforms all the other state-of-the-art algorithms. Taking into account of the image denoising results, our STLWSM has better robustness with much less PSNR changes compared to other competing approaches.

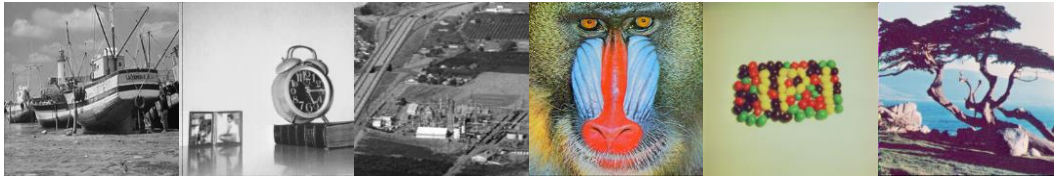

Fig. 10 Original images of size 512*512

Table 5 Images inpainting results of size 512*512

| Image | $\sigma_m$ | WNNM | SOLST | STROLLR | STLWSM |
|---|---|---|---|---|---|
| Boats | 15% | 57.88 | 56.51 | 56.88 | **58.05** |
| | 20% | 57.36 | 56.19 | 56.32 | **57.82** |
| | 30% | 56.57 | 55.76 | 55.87 | **57.32** |
| | 40% | 56.08 | 55.18 | 55.53 | **56.64** |
| | 50% | 55.86 | 54.79 | 55.05 | **55.63** |
| Clock | 15% | 54.39 | 53.85 | 53.91 | **55.12** |
| | 20% | 54.16 | 53.45 | 53.62 | **54.81** |
| | 30% | 53.99 | 53.76 | 53.94 | **55.52** |
| | 40% | 53.42 | 53.14 | 53.49 | **53.79** |
| | 50% | 52.15 | 52.14 | 52.50 | **52.71** |
| Factory | 15% | 59.11 | 58.18 | 58.26 | **59.68** |
| | 20% | 58.85 | 57.73 | 57.76 | **59.45** |
| | 30% | 58.26 | 56.16 | 56.35 | **58.98** |
| | 40% | 57.55 | 55.14 | 55.46 | **57.57** |
| | 50% | 56.86 | 54.49 | 55.11 | **56.66** |
| Baboon | 15% | 57.95 | 56.18 | 57.97 | **58.54** |
| | 20% | 57.25 | 55.85 | 56.95 | **57.94** |
| | 30% | 57.09 | 55.47 | 56.12 | **57.58** |
| | 40% | 56.56 | 54.95 | 55.27 | **57.07** |
| | 50% | 56.01 | 54.35 | 54.48 | **56.18** |
| Beans | 15% | 56.13 | 54.26 | 55.18 | **56.57** |
| | 20% | 55.85 | 54.19 | 54.79 | **56.19** |
| | 30% | 54.32 | 53.92 | 54.22 | **55.55** |
| | 40% | 53.61 | 53.14 | 53.29 | **54.74** |
| | 50% | 52.52 | 51.95 | 52.03 | **53.67** |
| Tree | 15% | 57.15 | 56.74 | 56.91 | **57.85** |
| | 20% | 57.08 | 56.34 | 56.66 | **57.64** |
| | 30% | 56.59 | 55.73 | 56.71 | **57.11** |
| | 40% | 54.73 | 54.67 | 54.71 | **55.26** |
| | 50% | 53.56 | 53.22 | 53.34 | **53.95** |

## Conclusions

In this paper, we have proposed a unified framework of image denoising using both knowledge from image domain and transform domain, namely Sparsity Transform Learning and Weighted Singular Values Minimization (STLWSM). Specifically, we learned the transform matrix for each group of patches with similar structure. After obtaining the optimized transform matrix and the sparse coefficient with an efficient optimization algorithm, we further restored the image patch groups through their low rank prior. By adopting STLWSM to all the groups, a denoised image can be reconstructed. For both gray images and color images, experimental results show that, the proposed model can achieve visible improvement in PSNR over other state-of-the-art approaches. Our efficient optimization algorithm also costs much less running time compared to the typical image domain based method. Note that while the pure transform learning methods run faster than STLWSM, they perform poorer with a large margin. To further improve the efficiency of our framework will be our main work in the near future.

## ACKNOWLEDGMENTS

The authors would like to thank the editor and the anonymous reviewers for their critical and constructive comments and suggestions. This work was supported in part by the National Key Research and Development Program of China under Grant 2018YFE0126100，National Science Fund of China under Grant Nos.51875524, 61873240 and 61602413, and the Natural Science Foundation of Zhejiang Province of China under Grant LY19F030016.